\documentclass[twocolumn,showpacs,showkeys,reprint,amsmath,amssymb,aps]{revtex4-1}

\usepackage{graphicx}
\usepackage{dcolumn}
\usepackage{bm}
\usepackage{tabularx}
\usepackage{amsmath}
\usepackage{xcolor}
\usepackage{float}

\emergencystretch=\maxdimen
\begin{document}

\title{Multigap nodeless superconductivity in the topological semimetal PdTe}

\author{Chengcheng Zhao,$^{1,2}$ Xiangqi Liu,$^{3}$ Jinjin Wang,$^{4}$ Chunqiang Xu,$^{5}$ Baomin Wang,$^{5}$ Wei Xia,$^{3,6}$ Zhenhai Yu,$^{3}$ Xiaobo Jin,$^{1}$ Xu Zhang,$^{1}$ Jing Wang,$^{1}$ Dongzhe Dai,$^{1}$ Chengpeng Tu,$^{1}$ Jiaying Nie,$^{1}$ Hanru Wang,$^{1}$ Yihan Jiao,$^{1}$ Daniel Duong,$^{7}$ Silu Huang,$^{7}$ Rongying Jin,$^{7}$ Zhu'an Xu,$^{4,\dag}$ Yanfeng Guo,$^{3,6,\ddag}$ Xiaofeng Xu$^{8,\sharp}$ and Shiyan Li$^{1,2,9,*}$}

\affiliation
 {$^1$State Key Laboratory of Surface Physics, Department of Physics, and Laboratory of Advanced Materials, Fudan University, Shanghai 200438, China\\
 $^2$Shanghai Research Center for Quantum Sciences, Shanghai 201315, China\\
 $^3$School of Physical Science and Technology, ShanghaiTech University, Shanghai 201210, China\\
 $^4$Zhejiang Province Key Laboratory of Quantum Technology and Device,Department of Physics, Zhejiang University, Hangzhou 310027, China\\
 $^5$School of Physical Science and Technology, Ningbo University, Ningbo 315211, China\\
 $^6$ShanghaiTech Laboratory for Topological Physics, Shanghai 201210, China\\
 $^7$SmartState Center for Experimental Nanoscale Physics, Department of Physics and Astronomy, University of South Carolina, Columbia, SC 29208, USA\\
 $^8$Department of Applied Physics, Zhejiang University of Technology, Hangzhou 310023, China\\
 $^9$Collaborative Innovation Center of Advanced Microstructures, Nanjing 210093, China
}

\date{\today}

\begin{abstract}
Recently PdTe was identified as a spin-orbit coupled topological Dirac semimetal and was claimed to exhibit both bulk-nodal and surface-nodeless superconducting gaps. Here we report the ultralow-temperature thermal conductivity measurements on PdTe single crystals with $T_c$ = 4.5 K to investigate its superconducting gap structure. It is found that the residual linear term $\kappa_0/T$ is negligible in zero magnetic field. Furthermore, the field dependence of $\kappa_0(H)/T$ exhibits an $\sf S$-shaped curve. These results suggest that PdTe has multiple nodeless superconducting gaps, which is at odds with the claimed bulk-nodal gap. The reason for the discrepancy is likely that previous angle-resolved photoemission spectroscopy measurements were only performed down to 2 K and cannot observe the smaller nodeless gap. The fully gapped superconducting state in PdTe is compatible with it being a topological superconductor candidate.
\end{abstract}

\maketitle

The exploration of superconductors and their superconducting mechanisms is an important frontier of condensed matter physics. High-temperature superconductors, unconventional superconductors, and topological superconductors are those novel ones to attract great attention. The pairing mechanism in unconventional superconductors is not phonon-mediated, and the superconducting gap usually manifests node (gap zero) \cite{Norman2011}. Superconductors with topological band structure bring the idea of topological superconductors (TSCs), which are a source of Majorana fermions and hold potential application in quantum computing \cite{RMP 2011}. In many cases, these three types of superconductors entangle with each other. For example, high-temperature iron-based superconductors are believed unconventional \cite{REV-Chen}, and some of them are topological \cite{CaKFeAs-nc,Science iron-based 2018,PhysRevX 2018,CPL 2019}. The excitation spectrum of a half-quantum vortex in a $p$-wave superconductor, theoretically predicted to feature a zero-energy Majorana fermion \cite{PRL-Ivanov}, has been exemplified by UTe$_2$ \cite{PRL UTe2 2019}, yet the experimental status remains inconclusive \cite{science UTe2 2021,JPSJ UTe2 2022,PRL UTe2 2023}. There are still considerable controversies about those topological superconductor candidates \cite{Science iron-based 2018,Zhang2019,Yin2015,Fang2019,Yuan2019,Sasaki2011}.

PdTe crystallizes in the NiAs-type hexagonal structure with the space group P6$_{3}$/mmc, as shown in Fig. 1(a), and has cell parameters $a$ = $b$ = 4.152 \AA\@ and $c$ = 5.671 \AA\@ \cite{strongly coupled}. It exhibits a superconducting transition temperature ($T_c$) of approximately 4.5 K \cite{strongly coupled,PNAS-PdTe}. While there remains a debate on whether PdTe is a strongly coupled superconductor or not, it is no doubt that PdTe is a type II superconductor \cite{strongly coupled,Type II}. During the exploration of iron-based high-temperature superconductors, PdTe was regarded as a distorted structure resulting from the sliding of anion and cation layers in iron-chalcogenides, and was investigated to get a clue for the pairing mechanism of iron-based superconductors \cite{Ekuma2013,Cao2016,Chen2016}. In contrast to iron-chalcogenides, PdTe exhibits robust covalent bonding, negligible electron correlation, and orbital non-degeneracy, thus tends to be a conventional superconductor \cite{Ekuma2013,Cao2016,Chen2016}.

However, recent angle-resolved photoemission spectroscopy (ARPES) study has revealed intriguing properties of PdTe, establishing it as a spin-orbit coupled Dirac semimetal with a topological Fermi arc across the Fermi surface \cite{PRL-PdTe}. More interestingly, ARPES measurements have identified bulk-nodal and surface-nodeless superconducting gaps in PdTe \cite{PRL-PdTe}. The subsequent specific heat measurements show that the electronic specific heat $C_e$ initially decreases in $T^3$ behavior, aligning with the characteristics of point nodes, and PdTe is suggested as an unconventional superconductor \cite{NSP-PdTe}. In this context, more experimental studies are highly desired to clarify the superconducting gap structure of PdTe.

Ultralow-temperature thermal transport represents a well-established bulk technique for investigating the energy gap of superconductors \cite{probe}. The existence of a finite residual linear term $\kappa_0/T$ in zero magnetic field is clear evidence for gap nodes \cite{probe}. Moreover, the field dependence of $\kappa_0/T$ may provide further information on the gap anisotropy, gap nodes, or multiple gaps \cite{probe}.

In this Letter, we present the ultralow-temperature thermal conductivity measurements on PdTe single crystal. A negligible $\kappa_0/T$ in zero field and an $\sf S$-shaped curve of $\kappa_0(H)/T$ are observed. These results suggest multiple nodeless superconducting gaps in PdTe. The reason for the discrepancy between our results and previous ARPES data is discussed.

Single crystals of PdTe were grown using the method described in Ref. \cite{strongly coupled}. The single crystals are stable in air. DC magnetization measurement was performed in a magnetic property measurement system (MPMS, Quantum Design). The specific heat was measured in a physical property measurement system (PPMS-9, Quantum Design) via the relaxation method. A piece of PdTe single crystal was cut into a rectangular shape with dimensions 1.39 $\times$ 0.129 $\times$ 0.071 mm$^3$. Four silver wires were attached to the sample with silver paint, which were used for both resistivity and thermal conductivity measurements. The low-temperature resistivity was measured in a $^3$He cryostat. The thermal conductivity was measured in a dilution refrigerator by using a standard four-wire steady-state method with two RuO$_2$ chip thermometers, calibrated $in$ $situ$ against a reference RuO$_2$ thermometer. To ensure a homogeneous field distribution in the sample, all fields for resistivity and thermal conductivity measurements were applied at a temperature above $T${$\rm_c$}.

\begin{figure}
\includegraphics[clip,width=8.5cm]{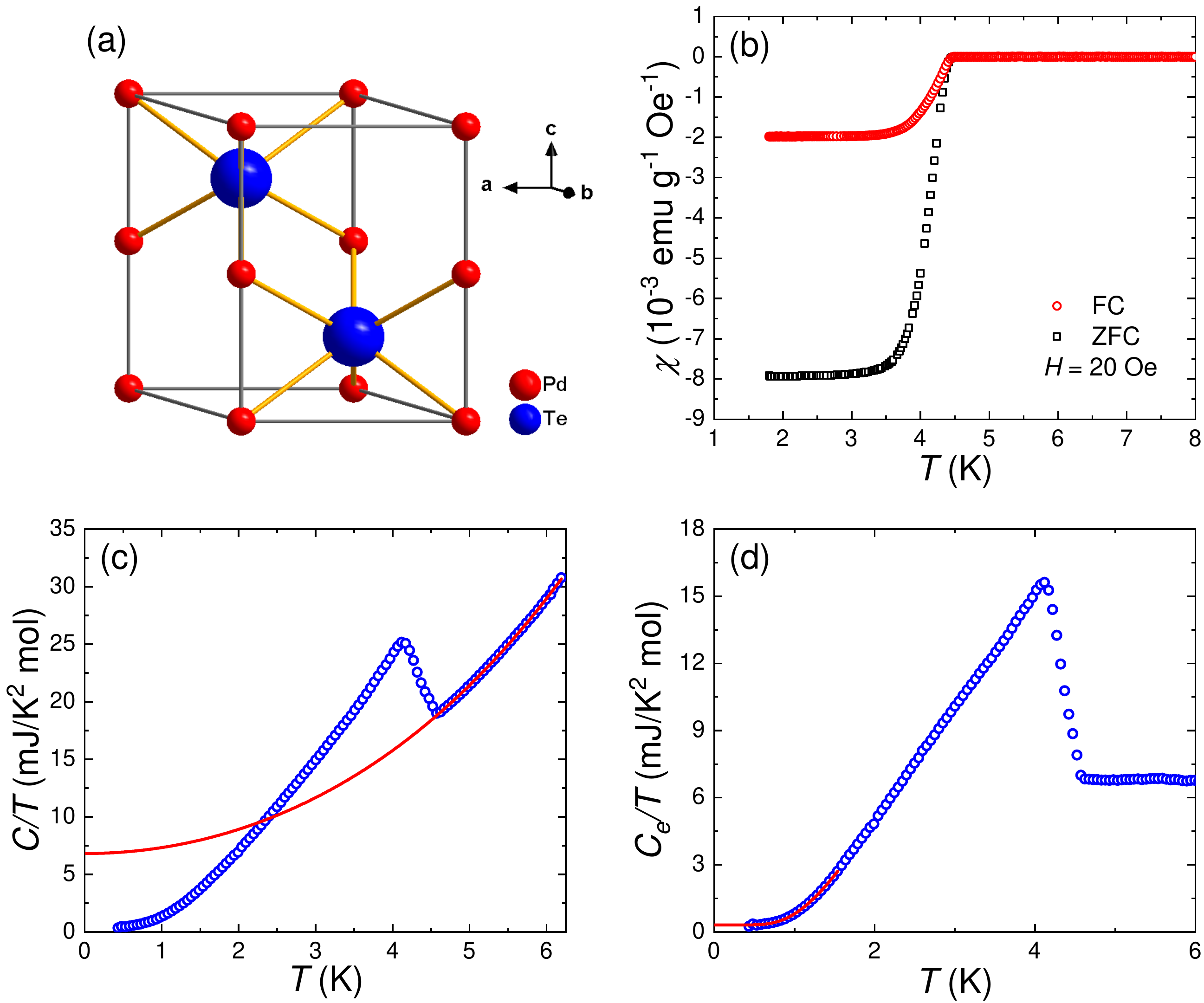}
\caption{(a) Crystal structure of PdTe. The Pd, Te atoms are presented as red and blue balls, respectively. (b) Low-temperature magnetization of a PdTe single crystal at $H$ = 20 Oe, with zero-field-cooled (ZFC) and field-cooled (FC) processes, respectively. (c) Low-temperature specific heat for PdTe, plotted as the $C/T$ vs $T$. The red line is a polynomial fit of the data between 4.62 and 6.19 K. (d) Electronic specific heat $C_e$ obtained from $C_e = C - \alpha T^{3} - \beta T^{5}$, plotted as the $C_e/T$ vs $T$. The red line represents a fit of the data below 1.5 K to $C_e/T =\gamma_r + A*exp(-\varDelta/kT)$.}
\end{figure}

\begin{figure}
\includegraphics[clip,width=7.2cm]{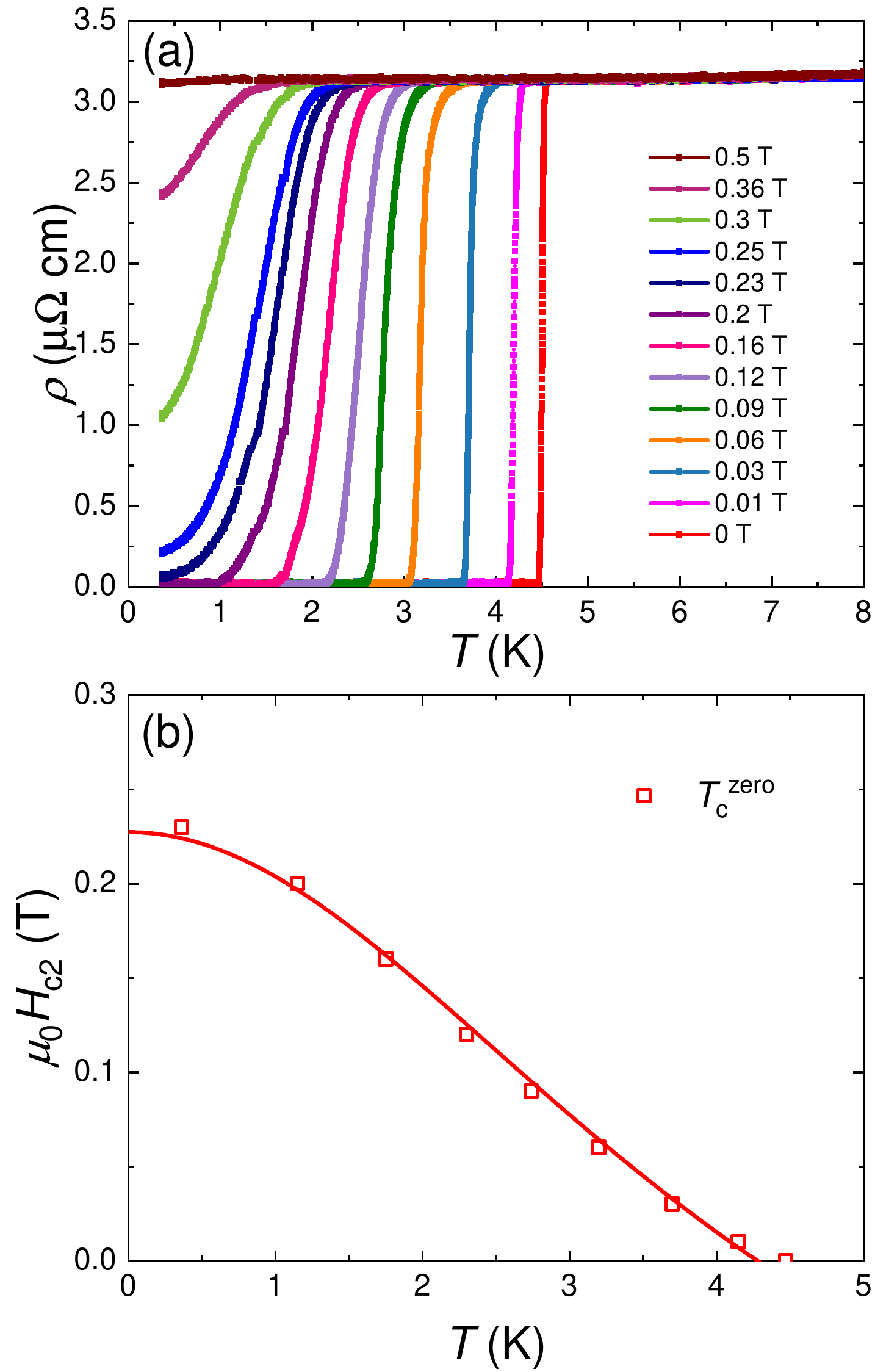}
\caption{(a) Low-temperature resistivity of a PdTe single crystal in magentic fields up to 0.5 T. (b) Temperature dependence of the upper critical field $\mu_0H{\rm_{c2}}$, extracted from the $T${$\rm_c \rm^{zero}$} values in panel (a). The red line is a fit to the Ginzburg-Landau equation $H{\rm_{c2}}(T) = H{\rm_{c2}}(0)[1 - (T/T{\rm_c})^{2}]/[1 + (T/T{\rm_c})^{2}]$, from which $\mu_0H{\rm_{c2}} \approx$ 0.23 T is roughly estimated.}
\end{figure}

The temperature dependence of the magnetic susceptibility from 1.8 to 8 K at 20 Oe with both zero-field-cooled (ZFC) and field-cooled (FC) processes is plotted in Fig. 1(b), showing the onset of the superconductivity at 4.5 K, in agreement with previous reports \cite{strongly coupled,PNAS-PdTe,PRL-PdTe,NSP-PdTe}. Figure 1(c) shows the temperature dependence of the specific heat $C(T)$ in the low-$T$ regime, plotted as $C/T$ vs $T$. Above $T_c$, from 4.62 to 6.19 K, the $C/T$ vs $T$ is well fitted by the formula $C/T = \gamma_n + \alpha T^{2} + \beta T^4$. The electronic specific heat coefficient $\gamma_n$, the phononic coefficient $\alpha$ and $\beta$ are determined to be 6.81 mJ K$^{-2}$ mol$^{-1}$, 0.51 mJ K$^{-4}$ mol$^{-1}$ and 0.0028 mJ K$^{-6}$ mol$^{-1}$, respectively. The electronic specific heat $C_e = C - \alpha T^{3} - \beta T^{5}$ is shown in Fig. 1(d), plotted as $C_e/T$ vs $T$. Below 1.5 K, $C_e/T$ shows an exponential $T$ dependence, and can be well fitted by the formula $C_e/T = \gamma_r + A*exp(-\varDelta/kT)$ with the energy gap $\varDelta$ = 0.387 meV and a finite residual linear term $\gamma_r$ = 0.31 mJ K$^{-2}$ mol$^{-1}$. The gap value is comparable to that obtained in Ref. \cite{strongly coupled,NSP-PdTe}, but our $\gamma_r$ is smaller.

Figure 2(a) depicts the low-temperature resistivity behavior of a PdTe single crystal in magnetic fields up to 0.5 T. The normal-state resistivity shows a very weak temperature dependence below 8 K. The superconducting transition is very sharp in zero field, and the transition is gradually suppressed to lower temperatures with increasing the field. In order to estimate the zero-temperature upper critical field $\mu_0H{\rm_{c2}}$(0), the temperature dependence of $\mu_0H{\rm_{c2}}$(T) is plotted in Fig. 2(b), defined by $\rho$ = 0 in Fig. 2(a). By fitting the data using the Ginzburg-Landau equation $H{\rm_{c2}}(T) = H{\rm_{c2}}(0)[1 - (T/T{\rm_c})^{2}]/[1 + (T/T{\rm_c})^{2}]$ \cite{PRL-Hc2,RMP-Hc2}, $\mu_0H{\rm_{c2}}(0) \approx$ 0.23 T is roughly estimated.

\begin{figure}
\includegraphics[clip,width=7.2cm]{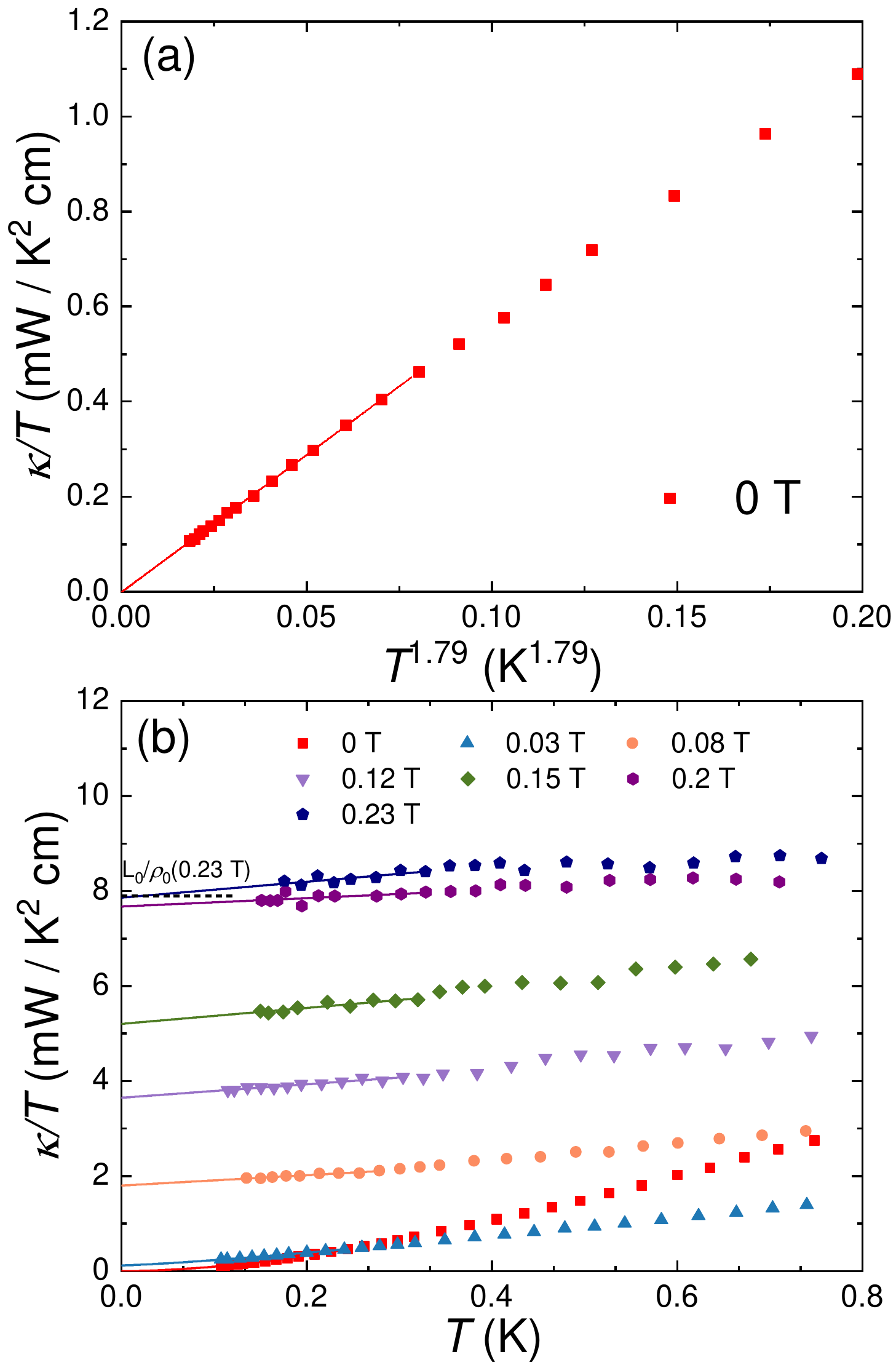}
\caption{(a) Temperature dependence of the thermal conductivity for the PdTe single crystal in zero field. The solid line represents a fit to $\kappa/T = a+bT^{\alpha - 1}$, which gives a negligible residual linear term $\kappa_0/T =$ -1.7 $\pm$ 0.5 $\mu$W K$^{-2}$ cm$^{-1}$. (b) Low-temperature thermal conductivity of the PdTe single crystal in magnetic fields up to 0.23 T. The dashed line is the normal-state Wiedemann-Franz law expectation $L_0/\rho_0$(0.23 T), with the Lorenz number $L_0 = 2.45 \times 10^{-8}$ W $\Omega$ K$^{-2}$ and $\rho_0$(0.23 T) = 3.11 $\mu \Omega$ cm.}
\end{figure}

Figure 3 shows the temperature dependence of the thermal conductivity of PdTe single crystal under zero and various magnetic fields, plotted as $\kappa/T$ versus $T$. The measured thermal conductivity contains two contributions, $\kappa = \kappa_e + \kappa_p$, which come from electrons and phonons, respectively. In order to separate the two contributions, the thermal conductivity in zero field is fitted to $\kappa/T = a + bT^{\alpha-1}$ \cite{Li2008,Sutherland2003}. The two terms $aT$ and $bT^\alpha$ represent contributions from electrons and phonons, respectively. The residual linear term $\kappa_0/T \equiv a$ is obtained by extrapolating $\kappa/T$ to $T$ = 0 K. The power $\alpha$ is typically between 2 and 3, due to specular reflections of phonons at the sample boundary \cite{Li2008,Sutherland2003}.

In zero field, the fitting results yield $\kappa_0/T = -1.7 \pm 0.5 $ $\mu$W K$^{-2}$ cm$^{-1}$ and $\alpha =$ 2.79 $\pm$ 0.04. Considering our experimental error bar of $\pm$ 5 $\mu$W K$^{-2}$ cm$^{-1}$, the $\kappa_0/T$ of PdTe in zero field is essentially zero. For $s$-wave nodeless superconductors, there are no fermionic quasiparticles to conduct heat as $T \rightarrow 0$ K, since the Fermi surface is entirely gapped \cite{Li2008,Sutherland2003}. Consequently, the absence of a residual linear term $\kappa_0/T$ should be observed, as seen in InBi and NbSe$_2$ \cite{Boaknin2003,PhysRevB.14.1916}. However, for nodal superconductors, a substantial $\kappa_0/T$ in zero field contributed by the nodal quasiparticles can be found \cite{probe}. For example, $\kappa_0/T$ of the overdoped ($T${$\rm_c$} = 15 K) $d$-wave cuprate superconductor Tl$_2$Ba$_2$CuO$_{6+\delta}$ (Tl-2201) is 1.41 mW K$^{-2}$ cm$^{-1}$, which is about $36\%$ of the normal-state value $\kappa{\rm_{N0}}/T$ \cite{Proust2002}. Therefore, the negligible $\kappa_0/T$ of PdTe strongly indicates a bulk nodeless superconducting gap.

Given $\mu_0H{\rm_{c2}}(0) \approx 0.23$ T, the fitting of the normal-state data in 0.23 T results in $\kappa_0/T = 7.86 \pm 0.25$ mW K$^{-2}$ cm$^{-1}$. This value of $\kappa_0/T$ agrees well with the normal-state Wiedemann-Franz law expectation $L_0/\rho_0$(0.23 T) = 7.88 mW K$^{-2}$ cm$^{-1}$, with the Lorenz number $L_0 = 2.45 \times 10^{-8}$ W $\Omega$ K$^{-2}$ and $\rho_0$(0.23 T) = 3.11 $\mu \Omega$ cm. The validation of the Wiedemann-Franz law in the normal state substantiates the reliability of our thermal conductivity measurements.

The field dependence of $\kappa_0/T$ can offer additional insights into the structure of the superconducting gap \cite{probe}. The normalized $\kappa_0/T$ as a function of $H/H{\rm_{c2}}$ for PdTe is plotted in Fig. 4, with $\kappa{\rm_{N0}}$(0.23 T)$/T$ = 7.86 mW K$^{-2}$ cm$^{-1}$ and $\mu_0H{\rm_{c2}}$(0) = 0.23 T. For comparison, similar data of the clean $s$-wave superconductor Nb \cite{Nb1970}, the dirty $s$-wave superconducting alloy InBi \cite{PhysRevB.14.1916}, the multiband $s$-wave superconductor NbSe$_2$ \cite{Boaknin2003} and PbTaSe$_2$ \cite{PbTaSe2}, and an overdoped $d$-wave cuprate superconductor Tl$_2$Ba$_2$CuO$_{6+\delta}$ (Tl-2201) \cite{Proust2002}, are also plotted.

In single-band clean $s$-wave superconductors, as well as in multiband $s$-wave superconductors as $T \rightarrow 0$ K, quasiparticles are localized within the vortex core, and conduction perpendicular to the magnetic field is predominantly attributed to the tunneling between adjacent vortices \cite{probe}. As the magnetic field increases, it becomes easier for quasiparticles to tunnel between vortex cores, leading to a slow exponential growth of $\kappa_0/T$ versus $H$, as is indeed observed in Nb \cite{Nb1970}. For dirty $s$-wave superconductor InBi, the response curve exhibits an exponential $H$ dependence at low magnetic fields, while manifests an approximately linear behavior as the magnetic field approaches $H{\rm_{c2}}$ \cite{PhysRevB.14.1916}. For nodal superconductor Tl-2201, a small field can yield a rapid growth due to the Volovik effect, and the low-field $\kappa_0(H)/T$ shows roughly a $\sqrt H$ dependence \cite{Proust2002}. In the case of multigap nodeless superconductors, the magnetic field dependence of $\kappa_0(H)/T$ relies on the ratio between the large and small gaps. For example, in the typical two-band superconductor NbSe$_2$, the ratio of different gap magnitudes is approximately 3 \cite{Boaknin2003}. A rapid rise in low fields can be attributed to the fast suppression of the smaller gap by the applied field.

\begin{figure}
\includegraphics[clip,width=7.5cm]{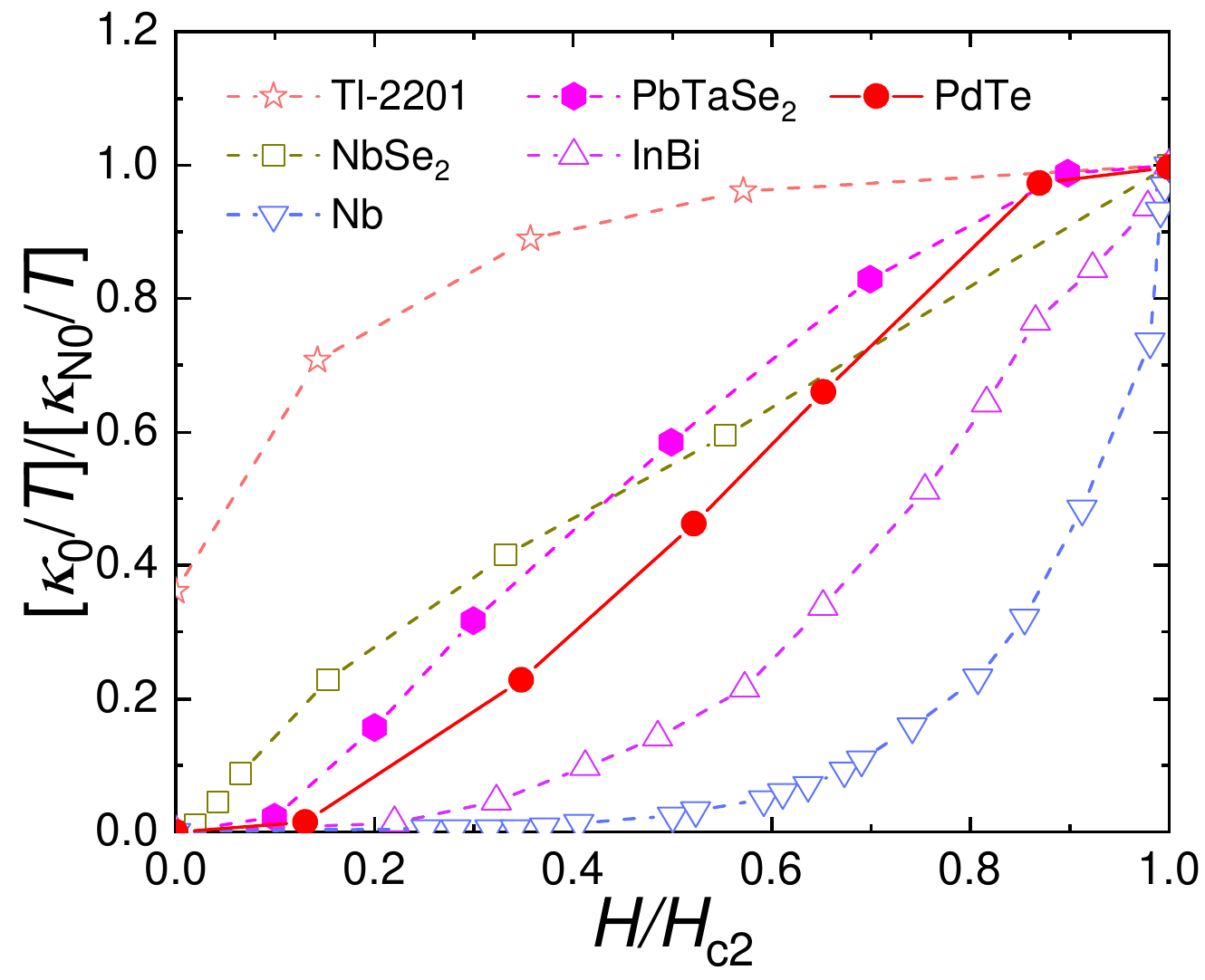}
\caption{Normalized residual linear term $\kappa_0/T$ of PdTe as a function of $H/H{\rm_{c2}}$. Similar data of the clean $s$-wave superconductor Nb \cite{Nb1970}, the dirty $s$-wave superconducting alloy InBi \cite{PhysRevB.14.1916}, the multiband $s$-wave superconductor NbSe$_2$ \cite{Boaknin2003} and PbTaSe$_2$ \cite{PbTaSe2}, and an overdoped $d$-wave cuprate superconductor Tl-2201 \cite{Proust2002} are shown for comparison.}
\end{figure}

From Fig. 4, the curve of the normalized $\kappa_0(H)/T$ for PdTe is very close to that of the multiband $s$-wave superconductor PbTaSe$_2$. Both of them are $\sf S$-shaped curves \cite{PbTaSe2}. Interestingly, nickel-based superconductors such as BaNi$_2$As$_2$, TlNi$_2$Se$_2$, and SrNi$_2$P$_2$ also exhibit $\sf S$-shaped field dependence of $\kappa_0(H)/T$ curves \cite{BaNi2As2,TlNi2Se2,SrNi2P2}. Such an $\sf S$-shaped $\kappa_0(H)/T$ curve is evidence for multiple nodeless superconducting gaps \cite{PbTaSe2,TlNi2Se2}. Indeed, the multiband electronic structure of PdTe has been demonstrated by density functional theory (DFT) calculations and de Hass-van Alphen (dHvA) oscillation experiments, and the superconducting specific heat can be well described with two nodeless energy gaps by using the two-band model to fit the data up to $T_c$ \cite{NSP-PdTe}.

While our ultralow-temperature thermal conductivity measurements clearly show multiple nodeless superconducting gaps in PdTe, previous ARPES measurements found that the bulk states near $\overline{Z}$ have a node in the superconducting state by exploring the bands close to the Brillouin zone boundary \cite{PRL-PdTe}. One most feasible explanation for this apparent discrepancy is that the ARPES measurements were only performed at 2 K (about 45$\%$ of the $T_c$), not low enough to observe the relatively smaller gap. In other words, the smallest gap among multiple superconducting gaps is still close at 2 K. This is supported by the exponential $T$ dependence of $C_e/T$ below 1.5 K, which suggests fully open superconducting gap at very low temperature. Note that in Ref. \cite{NSP-PdTe}, a residual linear term in specific heat $\gamma_r =$ 0.65 mJ mol$^{-1}$ K$^{-2}$ was obtained by the fitting, which is attributed to nodal quasiparticles or non-superconducting impure phase. In our specific heat data, the $\gamma_r$ is smaller (0.31 mJ mol$^{-1}$ K$^{-2}$). Since the absence of $\kappa_0/T$ in zero field in Fig. 3(a) indicates that there is no gap node, such a small $\gamma_r$ should be extrinsic.

Since the node previous ARPES measurements identified is a point node, another less possible explanation for the apparent discrepancy is that our heat flow is perpendicular to the nodal direction. In this case, we may not be able to observe the $\kappa_0/T$ in zero field contributed by the nodal quasiparticles. Limited by the small size of PdTe single crystals, during preparation of the rectangle sample, we are not able to determine the direction of the heat flow in it. The large surface of the reetangle sample is not a high-symmetry plane. Nevertheless, the chance for the heat flow to be perpendicular to the nodal direction should be very small. More experiments at low temperature (at least down to 10$\%$ of the $T_c$), such as scanning tunneling microscopy (STM), muon spin relaxation ($\mu$SR), and ARPES, are needed to further clarify the superconducting gap structure in PdTe. Finally, we would like to point out that both PdTe and PbTaSe$_2$ are superconductors with multiple nodeless gaps and topological band structure, which makes them candidates for natural topological superconductors.

In summary, we have examined the superconducting gap structure of the topological Dirac semimetal PdTe through ultralow-temperature thermal conductivity measurements. The negligible $\kappa_0/T$ at zero field and the $\sf S$-shaped field dependence of $\kappa_0(H)/T$ demonstrate multiple nodeless superconducting gaps in PdTe. This is in contrast to the bulk-nodal gap claimed by previous ARPES measurements at 2 K. In this sense, the superconductivity in PdTe may not be unconventional, but it is still a good candidate for topological superconductor.

We thank Rui Peng for helpful discussions. This work was supported by the Natural Science Foundation of China (Grant No. 12174064) and the Shanghai Municipal Science and Technology Major Project (Grant No. 2019SHZDZX01). Zhu'an Xu was supported by the Natural Science Foundation of China (Grant No. 12174334) and the National Key Projects for Research \& Development of China (Grant No. 2019YFA0308602).  Yanfeng Guo was supported by the open project of Beijing National Laboratory for Condensed Matter Physics (Grant No. ZBJ2106110017) and the Double First-Class Initiative Fund of ShanghaiTech University. Xiaofeng Xu was supported by the National Natural Science Foundation of China (Grants No. 12274369 and No. 11974061). Daniel Duong and Rongying Jin were supported by the grant No. DE-SC0024501 funded by the U.S. Department of Energy, Office of Science.

Chengcheng Zhao, Xiangqi Liu, Jinjin Wang contributed equally to this work.

\noindent $^\dag$ E-mail: zhuan$@$zju.edu.cn\\
\noindent $^\ddag$ E-mail: guoyf$@$shanghaitech.edu.cn\\
\noindent $^\sharp$ E-mail: xuxiaofeng$@$zjut.edu.cn\\
\noindent $^*$ E-mail: shiyan$\_$li$@$fudan.edu.cn

\end{document}